\documentstyle[epsfig]{mn}

\def\arcsecdot{\hbox{$.\!\!^{\prime\prime}$}}
\def\arcsec{\hbox{$^{\prime\prime}$}}

\def\ras{\hbox{$.\!\!^{\rm s}$}}

\def\arcmin{\hbox{$^\prime$}}

\def\speed{PSR B0906$-$49}
\def\vk{PSR J1105$-$6107}
\def\six{PSR B1610$-$50}
\def\four{PSR B1046$-$58}
\def\five{PSR B1055$-$52}
\def\svk{J1105$-$6107}
\def\ssix{B1610$-$50}
\def\sfour{B1046$-$58}
\def\sfive{B1055$-$52}
\newcommand{\lta}{\raisebox{-0.6ex}{$\,\stackrel
{\raisebox{-.2ex}{$\textstyle <$}}{\sim}\,$}}

\def\etal{et al. }
\def\edot{$\dot{E}$}
\def\tchar{$\tau_c$}
\def\gray{$\gamma$-ray}

\begin{document}

\title[A deep search for PWNe]{A deep search for pulsar wind nebulae using pulsar gating}
\author[B.W. Stappers et al. ]{B. W. Stappers$^{1}$\thanks{email :
bws@astro.uva.nl}, B. M. Gaensler$^{2,3,4,6}$, S. Johnston$^{5}$
\\
$^1$Sterrenkundig Instituut, Universiteit van
Amsterdam, Kruislaan 403,1098 SJ Amsterdam, The Netherlands.\\
$^2$Astrophysics Department, School of Physics A29, University of Sydney,
NSW 2006, Australia.\\
$^3$Australia Telescope National Facility, CSIRO, PO Box 76, Epping, NSW
2121, Australia.\\
$^4$Center for Space Research, Massachusetts Institute of Technology, 70
Vassar Street, Cambridge, MA 02139, United States of America.\\
$^5$Research Centre for Theoretical Astrophysics, University of Sydney, NSW
2006, Australia.\\
$^6$Hubble Fellow.\\}

\maketitle

\begin{abstract}
Using the Australia Telescope Compact Array (ATCA) we have imaged the fields
around five promising pulsar candidates to search for radio pulsar wind
nebulae (PWNe). We have used the ATCA in its pulsar gating mode; this
enables an image to be formed containing only off-pulse visibilities,
thereby dramatically improving the sensitivity to any underlying PWN. Data
from the Molonglo Observatory Synthesis Telescope were also used to provide
sensitivity on larger spatial scales. This survey found a faint new PWN
around PSR~B0906--49; here we report on non-detections of PWNe towards PSRs
\sfour, \sfive, \ssix\ and \svk.  Our radio observations of the field around
\five\ argue against previous claims of an extended X-ray and radio PWNe
associated with the pulsar. If these pulsars power unseen, compact radio
PWN, upper limits on the radio flux indicate that less than $10^{-6}$ of
their spin-down energy is used to power this emission. Alternatively \four\
and \six\ may have relativistic winds similar to other young pulsars and the
unseen PWN is resolved and fainter than our surface brightness sensitivity
threshold. We can then determine upper limits on the local ISM density of
$2.2\times10^{-3}$\,cm$^{-3}$ and $1\times10^{-2}$\,cm$^{-3}$,
respectively. Furthermore we constrain the spatial velocities of these
pulsars to be less than $\sim$450\,km\,s$^{-1}$ and thus rule out the
association of \six\ with SNR G332.4+00.1 (Kes 32). Strong limits on the
ratio of unpulsed to pulsed emission are also determined for three pulsars.
\end{abstract}


\begin{table*}
\begin{center}
\caption{Summary of the sample pulsars' parameters}
\begin{tabular}{rrrrrrrr}
\hline
\hline
\multicolumn{1}{c}{PSR}&
\multicolumn{1}{c}{l}&\multicolumn{1}{c}{b}&\multicolumn{1}{c}{Period}
& \multicolumn{1}{c}{$\tau_c$}& \multicolumn{1}{c}{\edot\ }    & \multicolumn{1}
{c}{DM} & \multicolumn{1}{c}{Distance} \\
             &  \multicolumn{1}{c}{($^{\circ}$)}  &
\multicolumn{1}{c}{($^{\circ}$)} &
\multicolumn{1}{c}{(ms)}  & \multicolumn{1}{c}{(kyrs)} &
\multicolumn{1}{c}{($10^{34}$ erg s$^{-1}$)} &  \multicolumn{1}{c}{(pc
cm$^{-3}$)} &  \multicolumn{1}{c}{(kpc)}   \\
\hline
B1046--58    &  287.43&  +0.58 &  124   &  20    &  200                 &  129  
   &   3.0    \\
B1055--52    &  285.98&  +6.65 &  197   &  535   &  3                   &   30  
   &   1.5    \\
J1105--6107  &  290.49&  --0.85 &   63   &  63    &  247                 &  270 
    &   7.0    \\
B1610--50    &  332.20&  +0.18 &  232   &   7    &  160                 &  582  
   &   7.2    \\
\hline
\end{tabular}
\label{char}
\end{center}
\end{table*}

\section{Introduction}

The majority of the spin-down energy \edot\ of a pulsar is carried away by a
relativistic wind which is a combination of Poynting and kinetic-energy flux
\cite{rg74,mic82}. The exact contributions to the wind from the magnetic
and particulate components is uncertain and has only been determined for the
Crab \cite{kc84} and PSR B1957+20 \cite{kpeh92}. In contrast, the
characteristic coherent pulsed radio emission represents only a small
fraction of the total energy liberated.

Particles emerging from a pulsar magnetosphere have a null pitch angle and
thus the relativistic wind is not directly observable. However, at a shock
interface these pitch angles are randomised and produce synchrotron emission
providing an excellent diagnostic of the wind (e.g. Frail \etal 1996). In
the Crab Nebula the pulsar's wind is confined in the high pressure
environment of the (invisible) surrounding supernova remnant (SNR) and a
compact ``plerion'' is observed \cite{wp78}. Outside this high-pressure
environment, the pulsar wind experiences the much lower pressure of the
general interstellar medium (ISM). This is expected to result in the
formation of very large ``ghost remnants'' \cite{bopr73}, an example of
which is yet to be discovered \cite{ccgm83}.  However, if the ram pressure
resulting from the motion of the pulsar relative to the ISM is high, the
interaction with the ambient medium may take the form of a bow-shock
nebula. These nebulae are collectively known as pulsar wind nebulae
(PWNe). Examples of plerions and bow-shock nebulae have been observed via
their emission at optical (H$\alpha$) (e.g. Kulkarni \& Hester
1988\nocite{kh88}; Bell, Bailes \& Bessell 1993\nocite{bbb93}), radio
(e.g. Frail \etal 1996\nocite{fggd96}), and X-ray (e.g. Wang, Li \& Begelman
1993\nocite{wlb93}) wavelengths. In general their radio emission is
distinguished by having a flat spectral index and significant linear
polarization. At present radio nebulae associated with pulsars are rare,
with only seven examples known.

Discovering more PWNe which emit at radio wavelengths is of great
interest. The spectral, polarization and energy properties of such nebulae
provide us with valuable information on the energy and particle composition
of the wind, the local ambient ISM density and the nebular magnetic
field. For a bow-shock nebula the cometary morphology and associated trail
can constrain both the transverse velocity of the pulsar and the direction
of that motion. This may be critical when considering possible SNR
associations where the pulsar is outside the remnant.

This paper reports on a search for radio PWNe around five southern pulsars.
It was partially inspired by the present sparsity of radio PWNe and also by
the recent VLA survey for PWNe carried out by Frail \& Scharringhausen
(1997, hereafter FS97)\nocite{fs97}. Their source list was made up of 35
candidates chosen for their high space velocity and/or high \edot. To avoid
contamination of a possible PWN by the pulsar itself they decided to observe
at 3~cm where the flatter spectrum of any PWN should cause it to dominate
its associated pulsar.  Unfortunately, however, FS97 detected no new nebulae
in their search.  As those radio PWN already observed have \edot\ $>
10^{35}$\,erg\,s$^{-1}$, while the majority of the pulsars searched had
\edot\ $< 10^{35}$\,erg\,s$^{-1}$, FS97 concluded that only young, energetic
pulsars produce radio-bright PWNe. However, their choice of observing
frequency and array configuration meant they could not have detected PWNe
larger than 20 arcsec, and also had poor sensitivity to low surface
brightness emission on smaller scales.  They also could not tell if
unresolved emission corresponded to compact PWNe or just to the pulsar
itself.

In this experiment, we have used pulsar gating on the Australia Telescope
Compact Array (ATCA) to remove the pulsed emission. This allows us to detect
compact PWNe ``hidden'' behind the pulsar itself.  It also enables us to
observe at lower frequencies and resolution, resulting in a significantly
more sensitive search. Our 20\,cm observations have a typical (1\,$\sigma$)
sensitivity of 70\,$\mu$Jy\,beam$^{-1}$ at $\sim$12 arcsec resolution,
compared to FS97's 40\,$\mu$Jy\,beam$^{-1}$ at $0.8$ arcsec. Hence, for a
typical PWN spectral index $\alpha = -0.3$ ($S_{\nu} \propto \nu^{\alpha}$),
we are able to detect a PWN of surface brightness \mbox{200 times fainter}
than could FS97. Our observations are also sensitive to PWNe on scales up to
$\sim$3 arcmin, $\sim$80 times the size of the largest sources which FS97
could detect. These observations were supplemented by Molonglo Observatory
Synthesis Telescope (MOST) observations at 36\,cm allowing us to probe
spatial scales up to 80 arcmin.

This search targetted five southern pulsars (\speed, \four, \five, \vk,
\six). A nebula associated with \speed\ was successfully detected and
is discussed in detail by Gaensler \etal (1998)\nocite{gsfj98b}. We
here report non-detections of PWNe towards the remaining candidates.
Candidate selection is discussed in Section 2, while in Section 3 we
discuss the observations and the pulsar-gating mode which we employed.
In Sections 4 and 5 we present our results and go on to consider the
corresponding implications.

\section{The Sample}

Because the ATCA is an East-West synthesis telescope, an image of each
pulsar requires a full 12-hour synthesis, and on a reasonable timescale it
is not possible to undertake a survey on the scale of that made by FS97.  We
therefore chose our sample of 5 pulsars based on a combination of their
characteristic age ($\tau_{c} = P/2\dot{P}$), \edot, and reported detections
of a PWN at other wavelengths. The characteristics of the four pulsars,
including their Galactic longitude, $l$, and latitude, $b$, rotational
period, characteristic age, \edot, dispersion measure, and the distance
implied by the dispersion measure \cite{tc93}, are given in Table
\ref{char}. We now consider each pulsar in detail.

\begin{table*}
\begin{center}
\caption{ATCA observations of PWN candidates.}
\begin{tabular}{rrrrrrrr}
\hline
\hline
\multicolumn{1}{c}{Source}&\multicolumn{1}{c}{Date}&\multicolumn{1}{c}{Secondary}&
\multicolumn{1}{c}{B$_{\rm
max}$}&\multicolumn{2}{c}{$\nu$\,(GHz)}&\multicolumn{1}{c}{$t_{res}$$^{\rm a}$} \\
\multicolumn{1}{c}{}&\multicolumn{1}{c}{}&\multicolumn{1}{c}{Calibrator}&\multicolumn{1}{c}{(m)}&\multicolumn{1}{c}{1}&\multicolumn{1}{c}{2}&\multicolumn{1}{c}{(ms)}\\
\hline
B1046--58  & 1997 May 13 & PKS B1215-457 & 5969 & 1.344 & 2.240 & 3.9\\
B1055--52  & 1997 May 14 & PKS B1215-457 & 5969 & 1.344 & 2.240 & 6.2\\
J1105--6107& 1995 June 1 & PKS B1036-697 & 750 & 1.376 & $\ldots$ & 7.9\\
J1105--6107& 1995 May 16 & PKS B1036-697 & 1485 & 1.376 & $\ldots$ & 7.9\\
B1610--50  & 1997 May 10 & MRC B1613-586 & 5969 & 1.344 & 2.496 & 8.0(7.2)\\
\hline
\end{tabular}
\label{obs}
\end{center}
(a) The values in parentheses corresponds to the effective time resolution at 13\,cm for
\six. For all other pulsars it does not differ from the effective time
resolution at 20\,cm. 
\end{table*}

\begin{figure}
\leavevmode\epsfig{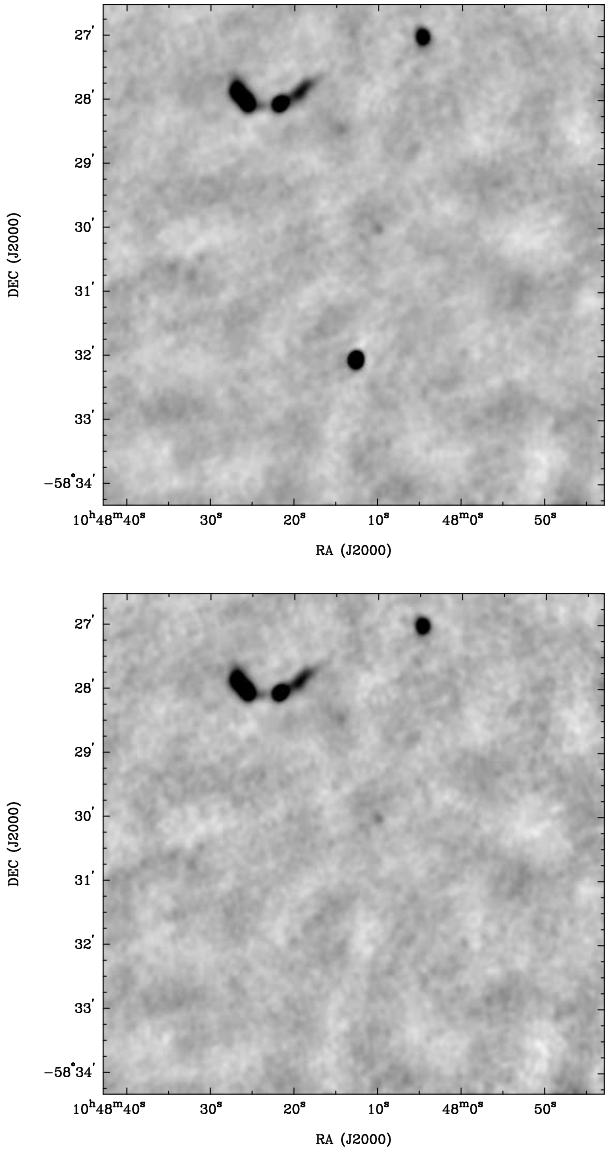}
\caption{Two {\tt CLEAN}ed images of the field around \four\ are
shown. Above all bins have been used; below only the off pulse bins are
included. The ease of identification of any non-pulsed emission are
immediately apparent. No PWN is discernible in the off-pulse image on any
spatial scale, a conclusion which can only be drawn when pulsar gating is
applied.}
\label{onoff}
\end{figure}

\subsection{\four}

\four\ was discovered in a high-frequency survey of the Galactic plane
\cite{jlm+92}. It is young, \tchar\ $\approx$ 20\,kyr, and has the 11th
highest \edot\ of all known pulsars. Despite its youthfulness it is not near
any known SNR. Its high \edot\ and relative proximity (2.5$\pm$0.5\,kpc,
Johnston \etal 1996\nocite{jkww96}) mean that it ranks in the top 10 pulsars
when considering the spin-down energy flux at the Earth. Thus, it is an
excellent candidate for having detectable high-energy emission (e.g. Fierro
1995\nocite{fie95}) and is indeed coincident with the {\em EGRET} source 2EG
J1049-5847 \cite{nab+96} and pulsations have probably been detected
\cite{klp+98}. There is also X-ray emission in the direction of the pulsar
which has been detected by both {\em ASCA} \cite{kt96,kts98a} and {\em ROSAT}
\cite{bt97}. The {\em ASCA} results suggest that this emission is extended
and may be from a pulsar-powered nebula.

\subsection{\five}

\five\ is not particularly young, energetic nor rapidly rotating.  However
it is one of only a small number of pulsars which are detected across the
whole electromagnetic spectrum and one of only seven pulsars known to pulse
in high-energy \gray s. It is a bright X-ray source \cite{ch83} with the
soft photons modulated at the pulse period \cite{of93} and is the pulsar
which converts the largest fraction of its spin-down energy into pulsed
\gray\ emission \cite{tbb+99}. Recent {\em HST} observations by Mignani,
Caraveo \& Bignami (1997)\nocite{mcb97} were successful in detecting the
optical counterpart. The optical emission is consistent with thermal
emission from the neutron star surface.

The {\em Einstein} detection of \five\ showed an apparently resolved source,
suggesting that the X-ray emission was from a PWN \cite{ch83}. However
{\em ROSAT} observations made by \"Ogelman \& Finley (1993) indicate no
extended structure. Subsequent {\em ASCA} observations in the 2--8\,keV
X-ray band show a clumpy 20 arcmin ring around the pulsar which has also
been interpreted as a PWN \cite{ssg+97}. Although not conclusive these
X-ray results suggest \five\ may be embedded in a PWN. Combi, Romero \&
Azc\'arate (1997)\nocite{cra97} have searched for an associated radio PWN
using a 30\,m dish at 20~cm. They detected a 66\arcmin$\times$126\arcmin\
radio source overlapping the pulsar's position and the proposed X-ray
nebula. Comparing this emission with a similar structure they see in the
408\,MHz (74\,cm) all-sky survey of Haslam \etal (1982)\nocite{hks+82} they
measured a non-thermal spectrum and suggested that the source may represent
synchrotron emission from a PWN powered by \five.

\subsection{\vk}

\vk\ is both young and energetic \cite{kbm+97}. Like \four\ its high
\edot\ suggests that it may be a \gray\ source and it is coincident with 2EG
J1103--6106. However Kaspi \etal (1998)\nocite{klp+98} find no evidence for
$\gamma$--ray pulsations attributable to \vk. This pulsar is also near the
SNR G290.1--00.8 and an association has been proposed by Kaspi \etal
(1997)\nocite{kbm+97}.  For the age and distance given in Table~\ref{char},
the association with G290.1--00.8 implies a projected velocity for the
pulsar of $v_{\rm t} \approx 650$\,km\,s$^{-1}$ in order to have reached its
present location well outside the SNR. X-ray observations with {\em ASCA}
indicate the presence of an unpulsed source at the pulsar's position which
is interpreted as a PWN \cite{gk98b}.

\subsection{\six}

\six\ is the third youngest known pulsar in the Galaxy. Its youth suggests
that it be associated with a SNR and indeed an association with the nearby
SNR G332.4+00.1 (Kes~32) has been claimed by Caraveo (1993)\nocite{car93}.
The pulsar is $\sim$12 arcmin from the center of the SNR, so an association
therefore requires a transverse velocity for the pulsar, $v_{\rm t} \approx
460\times d_{\rm kpc}$\,km\,s$^{-1}$, where $d_{\rm kpc}$ is the distance to
the pulsar in kpc. Its distance is quite uncertain, but taking the minimum
remnant/pulsar distance, $d_{\rm kpc} \sim 5$ (e.g. Johnston \etal
1995\nocite{jml+95}) gives a distinctly high implied velocity, $v_{\rm t} =
2300$\,km\,s$^{-1}$. Such a large velocity should result in a large ram
pressure with the ISM, and thus a radio-bright PWN with a cometary tail
indicating the pulsar's motion might be expected \cite{fk91}. As is
typical for a young pulsar, \six\ shows excessive timing noise and glitches,
and thus its position from pulse timing analysis is not well known
\cite{jml+95}. However an added advantage of the gated observations used
here is that the pulsar's position can be accurately determined.

\section{Observations and Data Reduction}

Observations made with the ATCA \cite{fbw92} are described in detail in
Table \ref{obs} where B$_{\rm max}$ is the maximum baseline and $\nu$ is the
central observing frequency. Observations were made simultaneously at
approximately 20~cm and 13~cm, except for \vk\ for which only 20~cm data
were obtained. A total bandwidth of 128\,MHz, divided into 32 channels, and
all Stokes parameters was recorded at both wavelengths. All observations
used the ATCA's pulsar gating mode, which divides the visibilities into a
maximum of 32 time bins per pulse period, with a minimum width of
$\sim$2.5\,ms. A polynomial description of the apparent pulse period is then
used to fold these bins online. Summed visibilities were then recorded every
40\,s. The more rapid rotation rate of \vk\ meant that only 8 bins were
used, while all 32 bins were recorded for the remaining pulsars. The
effective time resolution, $t_{res}$ for both frequencies is given in Table
\ref{obs}. The dispersion measure smearing inside each 4\,MHz channel was
less than or approximately equal to the sampling time in all cases except
for \six\ at 20~cm (both 20\,cm and 13\,cm resolutions for \six\ are
therefore given in Table \ref{obs}). For \four, \five\ and \six, the pulsar
was offset from the phase center by a few arcminutes.  Observations of \vk\
were centered on the SNR G290.1--01.8, so that the pulsar was approximately
22 arcmin away from the field center.

For all sources the flux density scale was determined by observations of PKS
B1934--638 using the revised flux density scale of Reynolds
(1994)\nocite{rey94}. The antenna gains were determined by observations,
every 40 minutes, of the secondary calibrators given in Table
\ref{obs}. Data reduction, including editing and calibrating, was carried
out using the {\tt MIRIAD} package and standard techniques
\cite{sk98}. Once calibrated, the data were dedispersed to the center
frequency of the observations using the dispersion measures (DMs) given in
Table \ref{char}. An image of each field was made, deconvolved using the
{\tt CLEAN} algorithm \cite{cla80}, and restored using a Gaussian beam
with dimensions appropriate for the diffraction limit (see $\theta_{min}$ in
Table \ref{limits}) and finally a correction for primary beam attenuation
was made.

Each pulsar was identified by subtracting, after dedispersion and
appropriate scaling, an image made from the first bin alone from an image
made using all bins. As the pulse was the only unique feature in any bin it
is the only source in the image\footnote{Noting that if the strongest
section of the pulse is in the first bin the pulsar will appear negative}. The
pulsar's position was determined by fitting a point source to the on-pulse
$u-v$ data, and the pulse profile was extracted from the entire data-set at
that position. The off-pulse bins were used to re-image the field and
generate an image which contained no pulsed emission. Any underlying PWN
emission can then be identified. Fig. \ref{onoff} compares the effect of
using all bins when observing \four\ (i.e.  equivalent to observing without
gating) with that of using only the off-pulse bins.

Our ATCA observations are only sensitive to a maximum spatial scale of
$\sim$3 arcmin. To search for more extended structures we used data from the
MOST \cite{rob91}, an East-West synthesis telescope operating at a fixed
wavelength of 36~cm. The MOST images all spatial scales between 43 arcsec
and $\sim$80 arcmin with a typical sensitivity (for a 12\,h synthesis) of
1\,mJy\,beam$^{-1}$, and is therefore particularly sensitive to large-scale
structure. These observations did not use pulsar gating. Data from the MOST
Galactic plane survey \cite{gcly99} were used for \four, \vk\ and \six,
while a separate observation was made of \five\ on 1998 Apr 4.

\begin{figure}
\leavevmode\epsfig{file=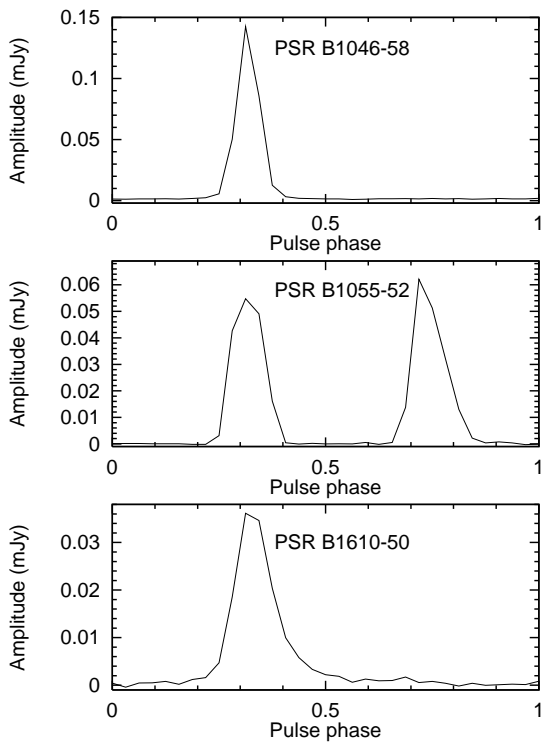, width=10cm,angle=-90}
\caption{The 20\,cm pulse profiles for the pulsars, as labelled, detected
in these observations. Profiles at 13\,cm looked similar. Thirty two time
bins were used and the time resolution is given in Table \ref{obs}. In all three
cases there are sufficient off-pulse bins to generate a high signal-to-noise
off-pulse map. Absolute timing was not carried out and the reported pulse
phase is therefore arbitrary.}
\label{profiles}
\end{figure}

\section{Results: Pulsed Emission}

Pulsations from \four, \five\ and \six\ were successfully detected in our
observations and their 20~cm pulse profiles are presented in Fig.
\ref{profiles}. These profiles, within the limited resolution, are
consistent with the known pulse shapes and similar profiles were obtained at
13\,cm. In all cases the pulsed emission can be clearly separated from any
off-pulse contribution. Flux densities and the interferometric positions for
all three detected pulsars are given in Table \ref{fluxes}.

\begin{table*}
\begin{center}
\caption{Observed parameters of the detected pulsars.$^{\rm a}$}
\begin{tabular}{rrrrrr}
\hline
\hline
\multicolumn{1}{c}{Name} & \multicolumn{2}{c}{Interferometric Position} 
& \multicolumn{3}{c}{Flux Density (mJy)} \\
&\multicolumn{1}{c}{R.A. (2000)} &
\multicolumn{1}{c}{Dec. (2000)} & 36\,cm & 20~cm & 13~cm \\
\hline
B1046--58 & 10:48:12.6(3)  & -58:32:03.75(1)  & 10(3)   & 9.4(5)  & 7.0(1) \\
B1055--52 & 10:57:58.9(9)  & -52:26:56.1(2)   & 15(2)   & 8.7(2)  & 3.5(1) \\
B1610--50 & 16:14:11.6(8)  & -50:48:01.90(9)  & $\dots$ & 4.7(2)  & 1.0(1) \\ 
\hline
\end{tabular}
\label{fluxes}
\end{center}
(a) The values in parentheses correspond to the 1\,$\sigma$ errors in the
final digit.\end{table*}

The flux of \four\ at 20~cm is consistent with that in the pulsar catalogue
\cite{tmlc95}, while \six\ has a 20~cm flux approximately twice that given
in the catalogue; no catalogue value for the 20~cm flux is given for
\five. A confusing thermal source, discussed below, is located near the
position of \vk\ and prevented its detection. The sensitivity limit for
detection was very close to the known 20~cm flux density of \vk, 1.8\,mJy
\cite{kbm+97}.

The position (Table \ref{fluxes}) of \five\ matches the timing position to
within the errors and our position for \four\ is only $\approx 2.3$ arcsec
different in declination. The position of \six\ is ill-determined from
timing and our improved position, RA (J2000) 16$^{\rm h}$14$^{\rm m}$11\ras6,
Dec (J2000) -50$^{\circ}$48\arcmin01\arcsecdot90, is approximately half an
arcminute away from the nominal timing position of Johnston \etal
(1995)\nocite{jml+95}.

\section{Results: Off-pulse Emission}

No off-pulse emission, either extended or unresolved, was detected from any
of the four pulsars discussed here. The 1\,$\sigma$ surface brightness
limits, $\sigma_{rms}$, for these non-detections are given in Table
\ref{limits}. These limits are determined by measuring the rms variation in
the off-pulse image in the vicinity of the pulsar. Also shown are the
minimum, $\theta_{min}$, and maximum, $\theta_{max}$, angular scales to
which we were sensitive at 20~cm. Two observations were made of \vk\ one of
which used a more compact configuration thereby increasing the angular size
to which we were sensitive.

\begin{table*}
\begin{center}
\caption{Limits on nebular and unpulsed flux densities.}
\begin{tabular}{rrrrrrrrrr}
\hline
\hline
\multicolumn{1}{c}{Source} & \multicolumn{3}{c}{$\sigma$\,(mJy/beam)} &
\multicolumn{1}{c}{$\theta_{\rm min}$} & \multicolumn{1}{c}{$\theta_{\rm
max}$} & \multicolumn{1}{c}{L$_{\rm R}$} & \multicolumn{1}{c}{L$_{\rm R}$/$\dot{\rm E}$} &
\multicolumn{2}{c}{$\sigma_{\rm off}$/S}\\
 & 36~cm & 20~cm & 13~cm & (\arcsec) & (\arcmin) & (erg s$^{-1}$) & ($10^{-6}$)& 20\,cm & 13\,cm \\
\hline
B1046--58  &  0.8  & 0.14 & 0.08 & 10 & 3.2 & 3$\times10^{29}$   & 0.15& 0.015 & 0.011\\
B1055--52  &  1.2  & 0.08 & 0.16 & 10 & 3.2 & 4.5$\times10^{28}$ & 1.5 & 0.009 & 0.05 \\
J1105--6107&  12   & 0.3  & $\ldots$  & 20 & 15.6& 2.5$\times10^{30}$ & 1.6 & $\ldots$ & $\ldots$ \\
B1610--50  &  5.0  & 0.2  & 0.16 & 12 & 3.2 & 3.5$\times10^{30}$ & 1.4 & 0.04  & 0.15 \\
\hline
\end{tabular}
\label{limits}
\end{center}
\end{table*}

The lack of any unpulsed emission in the vicinity of \four\ is starkly
apparent in Fig. \ref{onoff}. The 5\,$\sigma$ limits for any undetected
point-like nebula are 0.7\,mJy and 0.4\,mJy at 20 and 13~cm respectively.

\begin{figure}
\leavevmode\epsfig{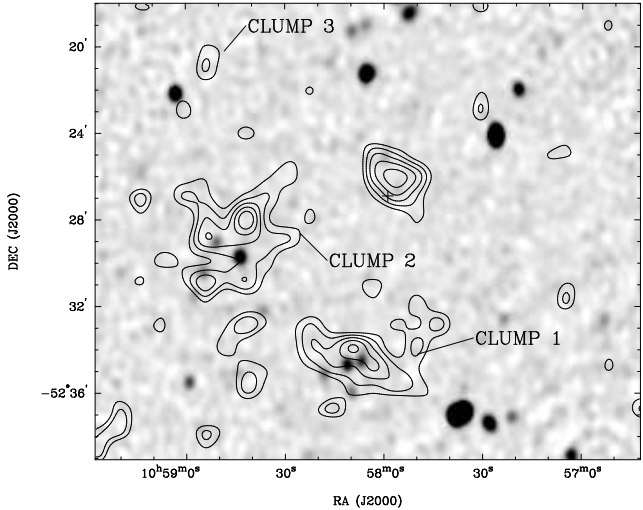}
\caption{Radio and X-ray observations of \five.  The grey-scale corresponds
to a 20\,cm off-pulse image, overlayed with X-ray contours from the {\em
ASCA} 0.5--2\,keV energy band.  The X-ray contours are at 50\%, 60\%, 70\%,
80\% and 90\% of the peak value. The position of the pulsar is labelled by
the cross and the three clumps of Shibata et al. (1997) are indicated. The
X-rays associated with the pulsar are clearly seen.  There is no evidence
for unpulsed radio emission at or around the position of the pulsar.}
\label{xrayneb}
\end{figure}

Fig. \ref{xrayneb} shows a 20\,cm off-pulse image of the field surrounding
\five\ with the {\em ASCA} X-ray contours in the energy band 0.5--2.0 keV
overlayed. The position of the pulsar is marked by the cross and the
contours corresponding to the X-ray emission from the pulsar and the three
``clumps'' \cite{ssg+97} are clearly seen. There is no indication of any
extended or point-like radio emission detected at the position of the
pulsar. Radio emission is seen at the position of at least two of the three
clumps, however it is clearly resolved into a number of discrete point
sources and does not appear to form part of a large extended structure.

\begin{figure*}
\leavevmode\epsfig{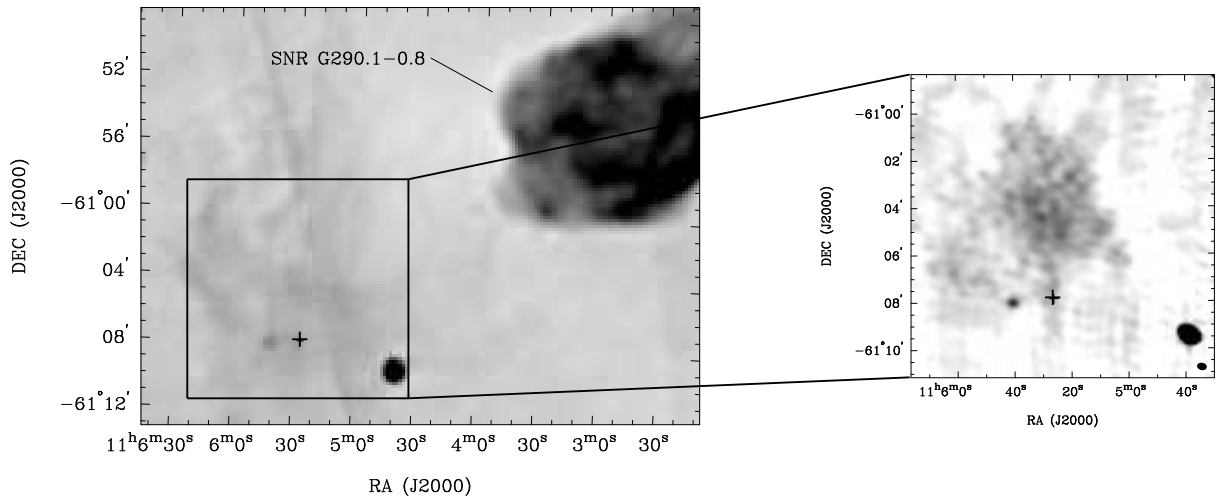}
\caption{Radio observations of \vk. Left: A 36~cm MOST image of the field,
including the SNR~G290.1--0.8. Right: A higher resolution, 20~cm, ATCA image
of the immediate region surrounding the pulsar. The ATCA synthesised beam is
shown at lower right. No pulsar gating has been applied to either image, and
the pulsar position is indicated by the cross.  The thermal emission
directly to the North of the pulsar can clearly be seen in the ATCA image.}
\label{vkfig}
\end{figure*}

The location of \vk\ relative to SNR~G290.1--0.8 is indicated by the cross
in the MOST 36\,cm image in the left hand panel of Fig.
\ref{vkfig}. Similarly the pulsar is marked in the higher resolution 20\,cm
ATCA image of the immediate region around the pulsar's location. Neither of
these images have been gated and the pulsar is undetected at both
frequencies. An extended source, whose center is some 4 arcmin to the North
of the pulsar, is clearly seen in the 20\,cm image and is also detected at
36\,cm. As this emission is diffuse and is coincident with an IRAS 60\,$\mu$m
source it is probably thermal and not associated with the pulsar. The pulsar
is un-detected but the 5\,$\sigma$ limit on any possible PWN radio emission
was only 1.5\,mJy at 20~cm.

The 36\,cm MOST image of the complicated field surrounding \six\ is shown in
Fig. \ref{full}. SNR~Kes 32 is seen to the north-east while SNR~G332.0+00.2
is to the south-west.  Using our ATCA data we also show a higher resolution
image of the region immediately surrounding the pulsar. After gating out the
pulsar's emission we detect no unpulsed emission except for a filled-center
clump located some 2 arcmin north of the pulsar which we designate
G332.23+00.2. No linear polarization was detected from G332.23+00.2 and we
measure flux densities of 0.20$\pm0.05$\,Jy (36~cm) and 0.20$\pm0.02$\,Jy
(20~cm), suggesting a flat spectral index, while it was largely resolved out
in our 13~cm observations.

\section{Discussion}

We have failed to detect any radio-bright PWNe emission over a wide range of
angular scales and down to good sensitivities around PSRs \sfour, \sfive,
\svk\ and \ssix. These non-detections have interesting implications for all
of these pulsars and we will consider them in detail below. Furthermore the
gating method employed here also allows us to put a strong limit on the
off-pulse emission from the pulsar itself. In Table \ref{limits} we give the
limits on the ratio of the off-pulse rms noise to the on-pulse flux
density. For \four\ and \five\ our limits are strong, and indicate that for
these young pulsars any unpulsed radio emission from the neutron star itself
is less than 2\% of the pulsed emission (c.f. Hankins \etal
1993)\nocite{hmnp93}.

For \vk\ and \six, we might have expected compact, ram-pressure confined
PWNe, resulting from the high velocities inferred through associations with
nearby SNRs. The emission from such nebulae may be located quite close to
the termination shock radius, $r_{\rm s}$. For example we can derive an
expression for the expected angular size of this shock (e.g. Cordes
1996)\nocite{cor96} near \six\ for a spherical wind;

\begin{equation}
\theta = \frac{0\arcsecdot15}{d_7v_{1000}}\sqrt{\frac{f}{n}},
\end{equation}
where $v = v_{1000}\times1000$\,km\,s$^{-1}$ is the pulsar's velocity, $d =
d_7\times7$\,kpc is the distance, $n$ is the density of the local ISM in
cm$^{-3}$ and $f$ is the efficiency of conversion of spin-down energy to
wind energy. Hence emission near $r_{\rm s}$ would be unresolved in our
observations under the assumption of high-velocity.

In fact, if we assume that all the pulsars in our sample are moving with a
typical pulsar velocity, 250\,km\,s$^{-1}$ (e.g. Hansen \& Phinney
1997\nocite{hp97}) and in ambient medium of density
$n\approx0.1$\,cm$^{-3}$, PWN around them will be unresolved in our 20~cm
data. Similarly to FS97, using our non-detections we can then determine an
upper limit on the radio luminosity, L$_{\rm R}$, for these compact
nebulae. Assuming a Crab-like spectral index (noting that the spectral index
of the PWN associated with \speed\ is steeper);

\begin{equation}
L_{\rm R} = 4.74\times10^{28}d^2_{kpc}S_{20}\,{\rm erg\,s}^{-1},
\label{rlum}
\end{equation}
where $d_{\rm kpc}$ is the distance to the nebulae and $S_{20}$ is the 20\,cm
flux limit. 

The L$_{\rm R}$ for each non-detection, based on the 5\,$\sigma$ limiting
flux density at 20\,cm, is given in Table \ref{limits}. Comparing these
maximum L$_{\rm R}$ values with the known \edot\ we find that the pulsars in
our sample have efficiencies (see Table \ref{limits}) which lie in or below
the peak of the distribution shown by FS97 at L$_{\rm
R}$/\edot$\sim10^{-6}$. Thus for an unresolved nebula the efficiencies are
well below those at which radio PWNe appear to be detected.

An alternative to this implied low efficiency for the lack of PWN could
simply be the fact that the PWN are resolved but are too faint to see. Both
\four\ and \six\ have similar ages, spin-down energies and surface magnetic
fields to PSR B1757--24 \cite{fggd96} and PSR B1853+01 \cite{fk91} which
are known to have associated PWN. If we therefore assume that these pulsars
all have similar wind characteristics and that conditions at the shock with
the ISM are the same, then the ratio of the resultant PWN's radio luminosity
to the spin-down energy should also be similar.

Using Equation \ref{rlum}, together with the distances and spin-down
energies given in Table \ref{char} and L$_{\rm R}$/\edot$\sim2\times10^{-4}$
(e.g. PSR B1757--24; FS97), we find that PWN associated with \four\ and \six\
should have S$_{20} = 1$ and 0.1\,Jy respectively. To reconcile these
expected flux densitys with the observed surface brightness limits these
nebulae must be larger than a certain radius. Assuming circular nebulae with
uniform surface brightness our 3\,$\sigma$ flux-density limits correspond to
PWNe of minimum radii 8 and 3\,pc respectively\footnote{The MOST offered the
best sensitivity limit for \four. Hence, assuming a Crab-like spectral
index, the 20\,cm flux limit was converted to a 36\,cm flux limit for
comparison with the MOST upper limit}.

Now consider that this minimum radius corresponds to the radius of a static
PWN \cite{aro93};

\begin{equation}
R_s = \left(\frac{\dot{E}}{4\pi\rho_0}\right)^{1/5}t^{3/5}{\rm cm},
\label{rstatic}
\end{equation}
where $\rho_0$ is the density of the surrounding medium in
g\,cm$^{-1}$\,s$^{-2}$ and $t$ is the age, we can derive a number density of
the surrounding medium, $n_0$. Around \four\
$n_0\lta2.2\times10^{-3}$\,cm$^{-3}$ and for \six\
$n_0\lta1.0\times10^{-2}$\,cm$^{-3}$, where we have assumed a purely
hydrogen ISM.

If instead we consider the case where the winds are in pressure balance at
the aforementioned radii due to ram pressure then we find that
$n_0v^2\lta0.4$\,cm$^{-3}$\,km$^{2}$\,s$^{-2}$ for \four\ and
$n_0v^2\lta2$\,cm$^{-3}$\,km$^{2}$\,s$^{-2}$ for \six. Thus for reasonable
pulsar velocities the implied density of the ISM is exceptionally low and we
conclude that these nebulae are probably not ram pressure confined. In this
case the pulsar velocity, $v$, cannot be significantly greater than the ISM
shock velocity and thus,
\begin{equation}
v \approx \dot{R}_S =
\left(\frac{3}{5}\right)\left(\frac{\dot{E}}{\rho_0t^2}\right)^{1/5}\,{\rm
cm\,s}^{-1}.
\label{time}
\end{equation}
Equating $t$ to the characteristic age of the pulsar and using
our limits on $n_0$ above, we obtain significant constraints on the maximum
space velocity of \four\ and \six\ of 480\,km\,s$^{-1}$ and
450\,km\,s$^{-1}$, respectively.

\subsection{\four}

\begin{figure*}
\leavevmode\epsfig{file=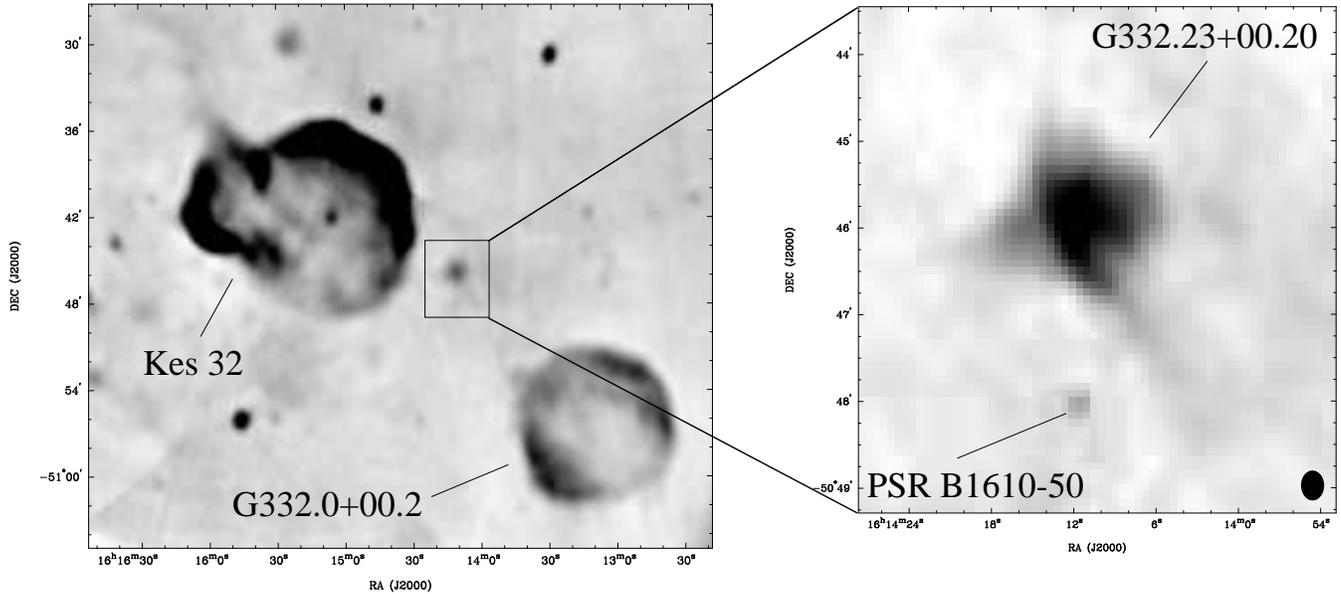, width=\linewidth}
\caption{Radio observations of \six. Left:  A 36~cm MOST image of
the field, including the SNRs Kes 32 and G332.0+00.2. Right: A higher
resolution, 20~cm, ATCA image of the immediate region surrounding the
pulsar. This image was formed by combining our ATCA data with other,
archival, observations of the region to give improved $u-v$ coverage
and sensitivity.  No pulsar gating has been applied to this image, and
the pulsar is indicated.  A knot of emission, G332.23+00.20, can be
seen to the north.  The ATCA synthesised beam is shown at lower right.}

\label{full}
\end{figure*}

\four\ is a young pulsar which we have shown is probably in a low density
region which is consistent with it having no associated SNR. Located near
the pulsar there is a 100\,pc expanding gas shell centered on a pair of OB
associations \cite{chty81}.  Johnston \etal (1996) speculate that this shell
may be related to the SN which formed \four. It is therefore probable that
despite being very energetic, \four\ dwells in an environment which
precludes the formation of a radio bright PWN.

\subsection{\five}

To properly compare our observations with the proposed extended nebula
detected by Combi \etal (1997) we smoothed our data to $\sim$24 arcmin, the
same resolution as their measurements. When smoothed, the distribution of
point sources shown in Fig.~\ref{xrayneb} almost exactly reproduces the
size and shape of the Combi \etal (1997) emission at 20~cm.  Furthermore,
the MOST data are sensitive to emission on scales as large as 80 arcmin, but
show no evidence for extended radio emission. Thus we conclude that the
source claimed as a PWN by Combi \etal (1997) is spurious, and is simply the
effect of confusing sources observed with a small dish.

The proposed X-ray PWN associated with \five\ is made of three clumps as
indicated in Fig. \ref{xrayneb}.  At least two of these three clumps
coincide with clusterings of radio point sources seen in our 20~cm
observations, and also with X-ray point sources seen in archival {\em
ROSAT}\ data \cite{bec98}. We thus think it more likely that the
{\em ASCA}\ clumps correspond to emission from unrelated background sources,
rather than to a PWN associated with \five.

To summarise, we argue that there is no evidence for either a radio or an
extended X-ray PWN around \five. It may be that \five\ is converting the
majority of its spin-down luminosity into high energy emission, rather than
into a relativistic wind. \five\ may emit as much as 21\% (assuming emission
into 1 steradian) of its spin-down energy at energies greater than 1\,eV
\cite{tbb+99} which is significantly more than for the other pulsars in our
sample.

\subsection{\vk}

Extensive thermal emission in the vicinity of \vk\ restricts the sensitivity
limits which we can place on any potential radio PWN. If this thermal source
is an H\,{\sc ii} region and lies along the line of sight to the pulsar then
this may mean that the pulsar distance derived from its dispersion measure
is significantly over-estimated.  Extrapolating the power-law spectrum of
the X-ray PWN ($\alpha =-0.8\pm0.4$; Gotthelf \& Kaspi 1998\nocite{gk98b})
through to radio wavelengths, we find that we expect a 20~cm flux density
for any radio PWN of $\sim$0.5~Jy. This corresponds to a 5\,$\sigma$ detection
of a $\sim$3 arcmin sized nebula in our 20~cm data.  The expected radio
flux falls below our detection threshold provided there is a break to a
Crab-like spectrum ($\alpha=-0.3$) at a frequency above 10$^{14}$\,Hz.

\subsection{\six}

As shown in Fig.~\ref{full}, the clump of emission G332.23+00.2 is
quite close to the pulsar.  While its filled-centre morphology and flat
spectrum are what might be expected from a PWN, the pulsar is well
outside this source and there is no morphological evidence to suggest
the two are associated. Rather, the non-detection of linear
polarization from G332.23+00.2 suggests that it is an unrelated thermal
region. The presence of strong IRAS 60\,$\mu$m emission nearby prevents a
clear determination of whether G332.23+00.2 also emits at this wavelength.  

The lack of a PWN associated with \six\ indicates that the pulsar is located
in a low-density environment and is moving with moderate velocity.  The
probability of a chance association along the line-of-sight to this pulsar
is high, as this complex region contains several SNRs and pulsars
(e.g. Johnston \etal 1995\nocite{jml+95}; Gaensler \& Johnston
1995\nocite{gj95b}).  Furthermore, unlike the case of PSR~B1757--24 and
G5.4-1.2 \cite{fk91}, there is no evidence of disruption of the SNR where
the pulsar would have passed through the shell.  From consideration of all
these points, we strongly argue that the association by Caraveo
(1993)\nocite{car93} of \six\ with Kes~32 is inconsistent with our data.

Thus, even though \six\ is a very young pulsar, there are no obvious SNRs
with which it might be associated. We do not consider this a problem,
however, we believe it to be located in a medium with density
$n_0\lta1.0\times10^{-2}$\,cm$^{-3}$ and since a large fraction of the ISM
is believed to consist of low density coronal gas (e.g. McKee \& Ostriker
1977\nocite{mo77}), SNRs expanding into such regions will be undetectable
(e.g. Kafatos \etal 1980\nocite{ksbg80}).

\section{Conclusions}

Using the excellent sensitivity to unresolved nebulae afforded by the gating
mode of the ATCA and the greater spatial sensitivity of the MOST we have
searched for PWNe around five pulsars. We successfully detected a new nebula
associated with \speed\ and failed to detect any extended or unresolved
nebulae associated with the remaining candidates. We also derived strong
limits on the unpulsed emission from PSRs \sfour, \sfive\ and \ssix.

These data argue strongly that the proposed ring-shaped X-ray nebula and
associated radio nebula around \five\ are in fact due to emission from
background sources. A lack of a nebula associated with \five\ combined with
its observed high energy characteristics indicates that it probably converts
most of its spin-down energy into high energy emission and not a
relativistic wind.  The lack of a nebular associated with \six\ clearly
shows that it is not moving sufficiently fast to be associated with
Kes\,32. \vk\ is located near the edge of a thermal region which will make
identification of any radio PWN difficult.

If the undetected nebulae around all our candidate pulsars are compact, then
these pulsars convert less than 10$^{-6}$ of their spin-down energy into
radio emission. Alternatively, assuming that \four\ and \six\ have similar
winds to PSR B1757--24 and PSR B1853+01 we derive a maximum ISM number
density for \four\ and \six\ of $n_0\lta2.2\times10^{-3}$\,cm$^{-3}$ and
$n_0\lta1\times10^{-2}$\,cm$^{-3}$, respectively. The low-density is
supported by the lack of any potential parental SNRs associated with these
young pulsars. We also constrain their spatial velocities to be less than
$\sim$450\,km\,$s^{-1}$. 

Our failure to detect nebulae around the young, and very energetic pulsars
\four\ and \six\ indicates that a high spin-down rate is insufficient to
generate a PWN. When combined with the very faint nebulae we did detect
associated with the somewhat less energetic PSR B0906--49 these results
highlight the role that the ISM plays in the production of such nebulae. The
sensitivity limits achieved here would be unattainable without the use of
pulsar gating and we now plan to employ this facility on the VLA to search a
larger sample of pulsars. Such a sample will not only provide vital data on
the wind parameters of a varied range of pulsars but will also supply
valuable information on the ISM.

\section*{Acknowledgments}

We would like to thank Dale Frail for valuable discussions during the
preparation of this work, Shinpei Shibata for providing the ASCA image of
\five\ and Dick Manchester for supplying the pulsar ephemerides. The
Australia Telescope is funded by the Commonwealth of Australia for operation
as a National Facility managed by CSIRO. The MOST is operated by the
University of Sydney with support from the Australian Research Council and
the Science Foundation for Physics within the University of Sydney. BMG
acknowledges the support of an Australian Postgraduate Award, and of NASA
through Hubble Fellowship grant HF-01107-01-98A awarded by the Space
Telescope Science Institute, which is operated by the Association of
Universities for Research in Astronomy, Inc., for NASA under contract NAS
5--26555.


\begin{thebibliography}{{Combi, Romero \& Azc\'{a}rate }{1997}}

\bibitem[\protect\citename{Arons}{1993}]{aro93}
Arons~J., 1993, ApJ, 408, 160

\bibitem[\protect\citename{Becker \& Tr\"{umper} }{1997}]{bt97}
Becker~W., Tr\"{umper}~J., 1997, AA, 326, 682

\bibitem[\protect\citename{Becker }{1998}]{bec98}
Becker~W., 1998, in The relationship between NSs and SNRs.
\newblock Osservatorio Astrofisico Di Arcetri, Elba,
  http://www.arcetri.astro.it/\~{ }elba98/

\bibitem[\protect\citename{Bell, Bailes \& Bessell }{1993}]{bbb93}
Bell~J.~F., Bailes~M., Bessell~M.~S., 1993, Nat, 364, 603

\bibitem[\protect\citename{Blandford {\rm et~al. }}{1973}]{bopr73}
Blandford~R.~D., Ostriker~J.~P., Pacini~F., Rees~M.~J., 1973, JA\&A, 23, 145

\bibitem[\protect\citename{Caraveo }{1993}]{car93}
Caraveo~P.~A., 1993, ApJ, 415, L111

\bibitem[\protect\citename{Cheng \& Helfand }{1983}]{ch83}
Cheng~A.~F., Helfand~D.~J., 1983, ApJ, 271, 271

\bibitem[\protect\citename{Clark }{1980}]{cla80}
Clark~B.~G., 1980, AA, 89, 377

\bibitem[\protect\citename{Cohen {\rm et~al. }}{1983}]{ccgm83}
Cohen~N.~L., Cotton~W.~D., Geldzahler~B.~J., Marcaide~J.~M., 1983, ApJ, 264,
  273

\bibitem[\protect\citename{Combi, Romero \& Azc\'{a}rate }{1997}]{cra97}
Combi~J.~A., Romero~G.~E., Azc\'{a}rate~I.~N., 1997, Astrophys. Space Sci.,
  250, 1

\bibitem[\protect\citename{Cordes }{1996}]{cor96}
Cordes~J.~M., 1996, in Johnston~S., Walker~M.~A., Bailes~M., eds, Pulsars:
  Problems and Progress, {IAU} Colloquium 160.
\newblock Astronomical Society of the Pacific, San Francisco, p.~393

\bibitem[\protect\citename{Cowie {\rm et~al. }}{1981}]{chty81}
Cowie~L.~L., Hu~E.~M., Taylor~W., York~D.~G., 1981, ApJ, 250, L25

\bibitem[\protect\citename{Fierro }{1995}]{fie95}
Fierro~J.~M., 1995, {\rm PhD thesis}, Stanford University

\bibitem[\protect\citename{Frail \& Kulkarni }{1991}]{fk91}
Frail~D.~A., Kulkarni~S.~R., 1991, Nat, 352, 785

\bibitem[\protect\citename{Frail \& Scharringhausen }{1997}]{fs97}
Frail~D.~A., Scharringhausen~B.~R., 1997, ApJ, 480, 364

\bibitem[\protect\citename{{Frail} {\rm et~al. }}{1996}]{fggd96}
{Frail}~D.~A., {Giacani}~E.~B., {Goss}~W.~M., {Dubner}~G., 1996, ApJ, 464, L165

\bibitem[\protect\citename{Frater, Brooks \& Whiteoak }{1992}]{fbw92}
Frater~R.~H., Brooks~J.~W., Whiteoak~J.~B., 1992, J. Electr. Electron. Eng.
  Aust., 12, 103

\bibitem[\protect\citename{Gaensler \& Johnston }{1995}]{gj95b}
Gaensler~B.~M., Johnston~S., 1995, Proc. Astr. Soc. Aust., 12, 76

\bibitem[\protect\citename{Gaensler {\rm et~al. }}{1998}]{gsfj98b}
Gaensler~B.~M., Stappers~B.~W., Frail~D.~A., Johnston~S., 1998, ApJ, 499, L69

\bibitem[\protect\citename{Gotthelf \& Kaspi }{1998}]{gk98b}
Gotthelf~E.~V., Kaspi~V.~M., 1998, ApJ, 497, L29

\bibitem[\protect\citename{Green {\rm et~al. }}{1999}]{gcly99}
Green~A.~J., Cram~L.~E., Large~M.~I., Ye~T.-S., 1999, ApJS, in press

\bibitem[\protect\citename{Hankins {\rm et~al. } }{1993}]{hmnp93}
Hankins~T.~H., Moffett~D.~A., Novikov~A., Popov~M., 1997, ApJ, 417, 735

\bibitem[\protect\citename{Hansen \& Phinney }{1997}]{hp97}
Hansen~B., Phinney~E.~S., 1997, MNRAS, 291, 569

\bibitem[\protect\citename{Haslam {\rm et~al. }}{1981}]{hks+82}
Haslam~C. G.~T., Klein~U., Salter~C.~J., Stoffel~H., Wilson~W.~E.,
  Cleary~M.~N., Cooke~D.~J., Thomasson~P., 1981, AA, 100, 209

\bibitem[\protect\citename{Johnston {\rm et~al. }}{1992}]{jlm+92}
Johnston~S., Lyne~A.~G., Manchester~R.~N., Kniffen~D.~A., D'Amico~N., Lim~J.,
  Ashworth~M., 1992, MNRAS, 255, 401

\bibitem[\protect\citename{Johnston {\rm et~al. }}{1995}]{jml+95}
Johnston~S., Manchester~R.~N., Lyne~A.~G., Kaspi~V.~M., D'Amico~N., 1995, AA,
  293, 795

\bibitem[\protect\citename{Johnston {\rm et~al. }}{1996}]{jkww96}
Johnston~S., Koribalski~B.~S., Weisberg~J., Wilson~W., 1996, MNRAS, 279, 661

\bibitem[\protect\citename{Kafatos {\rm et~al. }}{1980}]{ksbg80}
Kafatos~M., Sofia~S., Bruhweiler~F., Gull~S., 1980, ApJ, 242, 294

\bibitem[\protect\citename{Kaspi {\rm et~al. }}{1997}]{kbm+97}
Kaspi~V.~M., Bailes~M., Manchester~R.~N., Stappers~B.~W., Sandhu~J.~S.,
  Navarro~J., D'Amico~N., 1997, ApJ, 485, 820

\bibitem[\protect\citename{Kaspi {\rm et~al. }}{1998}]{klp+98}
Kaspi~V.~M., Lackey~J.~R., Pivovaroff~M.~J., Mattox~J.~R., Gotthelf~E.~V.,
  Manchester~R.~N., Bailes~M., Pace~R., 1998, in The relationship between NSs
  and SNRs.
\newblock Osservatorio Astrofisico Di Arcetri, Elba,
  http://www.arcetri.astro.it/\~{ }elba98/

\bibitem[\protect\citename{Kawai \& Tamura }{1996}]{kt96}
Kawai~N., Tamura~K., 1996, in Johnston~S., Walker~M.~A., Bailes~M., eds,
  Pulsars: Problems and Progress, {IAU} Colloquium 160.
\newblock Astronomical Society of the Pacific, San Francisco, p.~367

\bibitem[\protect\citename{Kawai, Tamura \& Saito }{1998}]{kts98a}
Kawai~N., Tamura~K., Saito~Y., 1998, Adv. Space Res., 21, 213

\bibitem[\protect\citename{Kennel \& Coroniti }{1984}]{kc84}
Kennel~C.~F., Coroniti~F.~V., 1984, ApJ, 283, 710

\bibitem[\protect\citename{Kulkarni \& Hester }{1988}]{kh88}
Kulkarni~S.~R., Hester~J.~J., 1988, Nat, 335, 801

\bibitem[\protect\citename{Kulkarni {\rm et~al. }}{1992}]{kpeh92}
Kulkarni~S.~R., Phinney~E.~S., Evans~C.~R., Hasinger~G., 1992, Nat, 359, 300

\bibitem[\protect\citename{McKee \& Ostriker }{1977}]{mo77}
McKee~C.~F., Ostriker~J.~P., 1977, ApJ, 218, 148

\bibitem[\protect\citename{Michel }{1982}]{mic82}
Michel~F.~C., 1982, Rev. Mod. Phys., 54, 1

\bibitem[\protect\citename{Mignani, Caraveo \& Bignami }{1997}]{mcb97}
Mignani~R., Caraveo~P.~A., Bignami~G.~F., 1997, ApJ, 474, L51

\bibitem[\protect\citename{Nel {\rm et~al. }}{1996}]{nab+96}
Nel~H.~I. {\rm et~al.}, 1996, ApJ, 465, 898

\bibitem[\protect\citename{\"Ogelman \& Finley }{1993}]{of93}
\"Ogelman~H., Finley~J.~P., 1993, ApJ, 413, L31

\bibitem[\protect\citename{Rees \& Gunn }{1974}]{rg74}
Rees~M.~J., Gunn~J.~E., 1974, MNRAS, 167, 1

\bibitem[\protect\citename{Reynolds }{1994}]{rey94}
Reynolds~J.~E., 1994, ATNF Technical Document Series, 39.3040

\bibitem[\protect\citename{Robertson }{1991}]{rob91}
Robertson~J.~G., 1991, Aust.\,J.\,Phys., 44, 729

\bibitem[\protect\citename{Sault \& Killeen }{1998}]{sk98}
Sault~R.~J., Killeen~N. E.~B., 1998, The Miriad User's Guide.
\newblock Australia Telescope National Facility, Sydney,
  (http://www.atnf.csiro.au/computing/software/miriad/)

\bibitem[\protect\citename{Shibata {\rm et~al. }}{1997}]{ssg+97}
Shibata~S. {\rm et~al.}, 1997, ApJ, 483, 843

\bibitem[\protect\citename{Taylor \& Cordes }{1993}]{tc93}
Taylor~J.~H., Cordes~J.~M., 1993, ApJ, 411, 674

\bibitem[\protect\citename{Taylor {\rm et~al. }}{1995}]{tmlc95}
Taylor~J.~H., Manchester~R.~N., Lyne~A.~G., Camilo~F.
\newblock 1995.
\newblock Unpublished (available at ftp://pulsar.princeton.edu/pub/catalog)

\bibitem[\protect\citename{Thompson {\rm et~al. }}{1999}]{tbb+99}
Thompson~D.~J. {\rm et~al.}, 1999, ApJ, accepted, astro-ph/9811219

\bibitem[\protect\citename{Wang, Li \& Begelman }{1993}]{wlb93}
Wang~Q.~D., Li~Z.-Y., Begelman~M.~C., 1993, Nat, 364, 127

\bibitem[\protect\citename{Weiler \& Panagia }{1978}]{wp78}
Weiler~K.~W., Panagia~N., 1978, AA, 70, 419

\end{thebibliography}
\end{document}